# Experimental study of plasma window[1*]

SHI Ben-Liang(史本良), HUANG Sheng(黄胜) , ZHU Kun(朱昆)[1)], LU Yuan-Rong(陆元荣)
State Key Laboratory of Nuclear Physics and Technology, Department of Physics, Peking University, Beijing 100871, China

**Abstract:** Plasma window is an advanced apparatus which can work as the interface between vacuum and high pressure region. It can be used in many applications which need atmosphere-vacuum interface, such as gas target, electron beam welding, synchrotron radiation and spallation neutron source. A test bench of plasma window is constructed in Peking University. A series of experiments and corresponding parameter measurements have been presented in this article. The experiment result indicates the feasibility of such a facility acting as an interface between vacuum and high pressure region.

1. Introduction

Since initially described by Sir Humphry Davy at the beginning of the nineteenth century, electric arc has been widely applied to welding, cutting, metallurgy and various kinds of lamp in the industrial field [1]. In laboratory study and realistic manufacturing of spacecraft thrusters, the electric arc has also been highly involved [2]. Due to the high flux of plasma, some devices have been built up for studying the interaction of plasma and material in International Thermonuclear Experimental Reactor(ITER) plan recently [3, 4]. In 1995, the concept of plasma window was originally proposed by Ady Hershcovitch from BNL [5]. The following experiments showed plasma window can separate a vacuum of $7.6 \times 10^{-6}$ torr from atmosphere for argon [5]. Transmission of electrons, x-ray and ions has been proven successfully by some experiments [6]. High-quality electron beam welding with plasma shielding was achieved [7]. The basic characteristics of the plasma window have been described in some papers [5, 8].

The conventional metallic foil window used in gas target, such as aluminium window, molybdenum window and Havar window(a kind of colt-based non-magnetic alloy), cannot be very thin because of strength requirements for pressure differential. Therefore a foil window can cause beam energy loss, especially in the case of low-energy ion beams. Besides, the beam intensity is limited under the consideration of cooling problem and lifetime of the window. Compared with the foil window, the plasma window can sustain high-current ion beam almost without energy loss. In addition, the invulnerable feature of plasma window can enable the whole system working stably in the long run. For further application, a deuterium gas target with plasma window has been built to generate mono-energetic fast neutron[9]. Plasma window can also work as an interface which separates the lead-bismuth eutectic target from the vacuum of a particle accelerator in ADS (Accelerator-driven Subcritical Nuclear Energy System). The high pressure region generated by plasma window is useful for controlling the free surface of liquid metal[10].

2. plasma window device

* Supported by National Natural Science Foundation of China (10805003 and 91026012)
[1)] E-mail: zhukun@pku.edu.cn

Plasma window is a channel filled with plasma generated by stable DC discharge arc. It consists of three cathodes, an anode and a stack of six cooling copper plates. The thickness of plates is 9mm, and thin plate design makes it easy to ignite because of improved electric field. The six plates are insulated from each other by boron nitride spacers, which have remarkable thermal stability and conductivity among available insulating materials. The cylindrical apertures in the center of each copper plate and BN spacers form a DC discharge space. In this experiment, copper plates with different diameter are used: one is 3mm and the other is 6mm. Water cooling is critical for the stable operation of arc, so all components which are close to the arc have well-designed water cooling channels. Compared with the common free-burning arc, such wall-stabilized arc can guarantee a higher voltage with a smaller current. Therefore arc power can be increased while the life span of cathodes can be expanded largely [11]. The schematic is shown in FIG.1. The similar design was also used in the research on the physical properties of plasma [12].

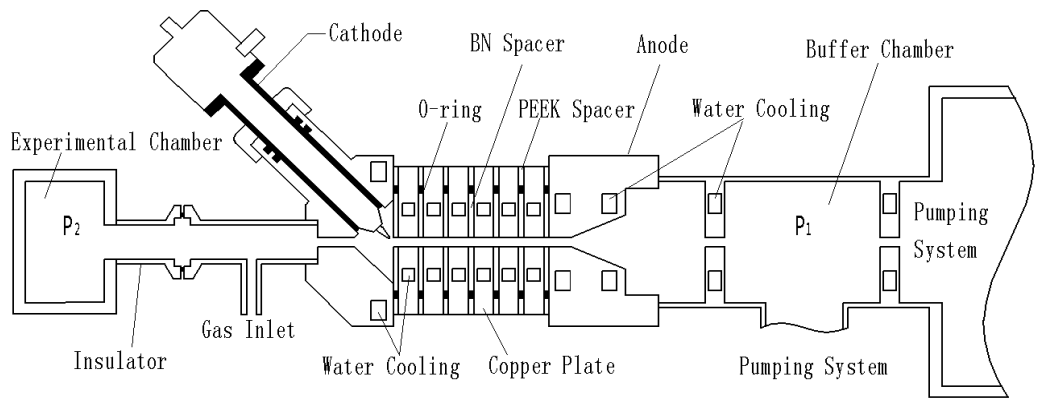

FIG. 1. Schematic diagram of plasma window

The separation function of plasma window mainly depends on pressure equalization effect [5]. According to the equation of state for ideal gas:

$$p = nkT,$$

considering the average temperature is 12,000K within the arc and 300K in the experimental chamber, the density in arc is 1/40 of the gas density of the experimental chamber [5]. Dynamic viscosity effect is another contributive factor. It is known that the pressure will drop while a laminar gas flowing through a tube, which can be roughly described by the Poiseulle's equation for compressible ideal gas:

$$p_2^2 - p_1^2 = \frac{16}{\pi}\eta\frac{l}{r^4}NRT,$$

where $r$ and $l$ are the tube radius and length, $N$ represents the mole number of gas flow rate, $\eta$ is the dynamic viscosity coefficient. The viscosity of plasma gas depends highly on temperature, but varies little with pressure. According the data of Ref.13, the viscosity of argon plasma is 2.42×10$^{-4}$, 10 times higher than that of argon in the room temperature. These properties make the plasma window form a pressure differential between two sides. If pumping system is arranged properly, the pressure differential can be very high.

3. Experiment results of plasma window

The experiment system comprises a stabilized current power supply, a differential pump system with combination of roots pump and claw-type pump, measurement instruments and plasma window. The photograph of plasma window is showed in FIG.2. The stabilized current supply provides a direct current output from 10A to 80A, a DC voltage output from 40V to 400V, and an ultimate power output up to 30KW. A pressure sensor is mounted on the side of the experiment chamber with high pressure. On the other side of plasma window is a buffer chamber connected with pump and vacuum gauge. Integrated with a high frequency triggering system, the power supply provides an alternating output voltage more than 6000V for ignition. In all previous designs, many additional devices are used, including Tesla coil or other setups for ignition, LC filters to reduce the electromagnetic noise and high-power resistors with low Ohm value to further stabilize the current.

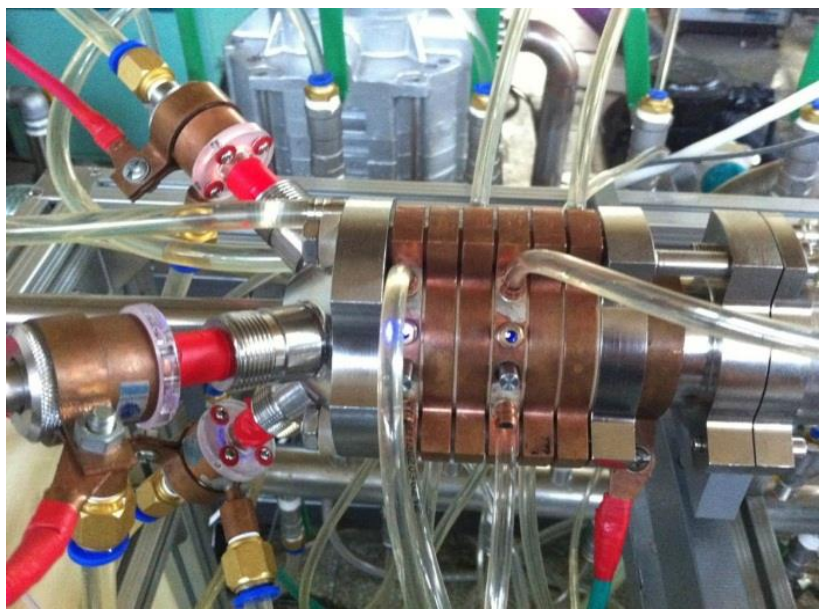

FIG. 2 Photograph of Plasma Window

A series of measurements have been done to test the performance of the entire system. The plasma window operates under various powers and gas flow rates with argon. In the experiment, only the gas flow rate and the discharge current can be adjusted. The power consumed by the arc discharge is related to the pressure $P_2$ which can influence the function of the mass flow controller. By adjusting gas flow rate and current simultaneously, $P_1$ can be set at a constant value 60pa. In this way the data of $P_2$ have comparability, so as the pressure reduction factor(pressure reduction factor is defined as $P_2/P_1$). .

At first, different pressure data without discharge are measured by adjusting the flow rate. The relationship of $P_1$ and $P_2$ is shown in the Fig.3. For the 6mm aperture, the pressure of $P_1$=60pa in the buffer chamber corresponds to $P_2$=2.6kpa in the experimental chamber. The pressure reduction factor is 43. For the 3mm aperture, the factor is about 100 and $P_2$=5.3kpa. The relationship between $P_1$ and $P_2$ is showed in FIG.3. The pressure sensor is used for the measurement of $P_2$, the error is less than 10% around 0.1atm. For the measurement of $P_1$, the error of quartz vacuum gauge is less than 30% in the range of 10Pa to 100Pa. All the pressure data are the average of 20 records.

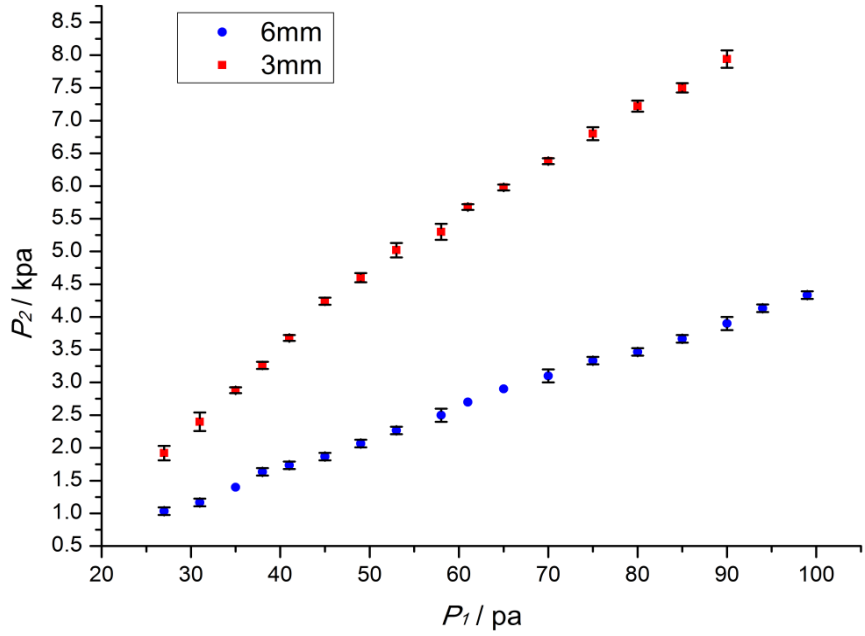

FIG. 3. $P_1$ and $P_2$ with different gas flow rate

The pressure $P_2$ under different power is shown in the FIG.4. A factor up to 200 has been achieved for the 6mm aperture with stable arc discharge, and 800 for the 3mm aperture. Although the reduction factor decreases 3 times from 6mm to 3mm at the same power supply, it increases with power. A further improvement space for a larger diameter can be speculated if the higher power supply is used.

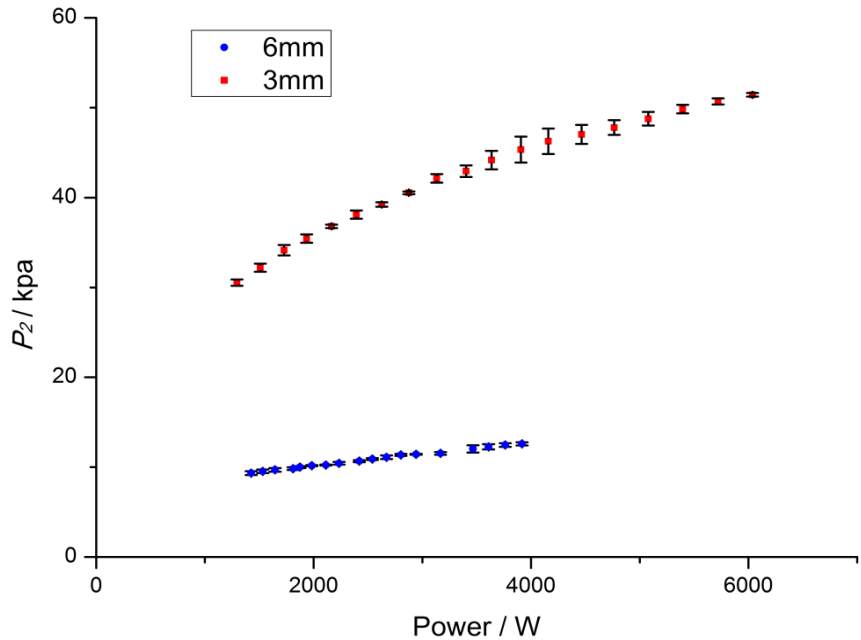

FIG 4 $P_2$ under different power

A typical potential distribution of copper plates with 50A discharge current is shown in Table.1. It is reasonable to assume that the plasma in the middle of the discharge column is fully developed, so the average electric field intensity is calculated based on the voltage drop between the second and fifth cooling plates. For the voltage drop between the copper plate and the anode, there is a fluctuation of 0.1 or 0.2V. And for the voltage drop between anode and the cathode, the fluctuation is less than 0.4V. The average plasma conductivity can be calculated by: $\sigma = I / \pi r^2 \overline{E}$ [8]. It is $30.6 \Omega^{-1} cm^{-1}$ for 3mm aperture, and $21.2 \Omega^{-1} cm^{-1}$ for 6mm. In the Ref.8, $\overline{E}$ is the average electric field between the cathodes and the first copper plate, which is inappropriate because of the existence of steep potential fall near the cathodes. According to the data of Ref.13, the average temperature is estimated to be 1,0100K for 3mm and 8900K for 6mm.

TABLE I. The copper plate potential distribution of different aperture diameters: 3mm & 6mm

|  | 1st | 2nd | 3rd | 4th | 5th | 6th | Anode | $P_1$(Pa) | $P_2$(kpa) |
|---|---|---|---|---|---|---|---|---|---|
| **3mm** | 14.2 | 36.1 | 52.2 | 68.3 | 84.6 | 100.2 | 114.4 | 60 | 46.3 |
| **6mm** | 11.6 | 20.8 | 26.5 | 32.2 | 38.3 | 43.8 | 53.4 | 60 | 11 |

4. Conclusions

In summary, a 6mm diameter of discharge channel and a 6.1cm total length are the largest in the similar experiments. A bore diameter as large as possible is required in the application of ion beam field. The experimental data presented in this paper demonstrate the feasibility of the plasma window.

Only one pump system is enabled in this experiment. Its effective pumping rate under 100pa is less than 1L/s. In other experiments, there are several chambers separately connected independent pumps. It is reasonable to conclude that a higher pumping rate can secure a better pressure differential. The following experiments will focus on the optimization of the operating parameters and larger diameter. Besides, further structure modifications are proposed. In the experiments, water cooling is found to be a very crucial factor, especially for the buffer chamber due to the contact with anode jet.